# Local-time effect on small space-time scale.

V.A. Panchelyuga[1], V.A. Kolombet[1], M.S. Panchelyuga[1], S.E. Shnoll[1,2]

*Lomonosov Moscow State University* (1)
*Institute of Theoretical and Experimental Biophysics of RAS* (2)

*shnoll@iteb.ru, panvic333@yahoo.com*

The paper presents an investigation of local-time effect - one of the manifestations of macroscopic fluctuations phenomena. Was shown the existence of the named effect for longitudinal distance between locations of measurements up to 500 meters. Also a structure of intervals distribution in neighborhood of local-time peak was studied and splitting of the peak was found out. Obtained results lead to conclusion about sharp anisotropy of space-time.

## 1. Introduction.

Present work was carried out as further investigations of macroscopic fluctuations phenomena [1 – 4]. The local time effect, which is the main subject of this paper, is synchronous in local time appearance of pairs of histograms with similar fine structure constructed on the base of measurements of fluctuations in processes of different nature fulfilled in different geographical locations. The effect points out on the dependence of fine structure of the histograms on the Earth rotations around its axis and around the Sun. The existence of local time effect was studied for different distances between places of measurement from hundred kilometers up to highest possible on the Earth distances (~ 15000 km). The goal of the present work is the investigations of the existence of the local time effect for relatively small distances between places of measurements.

      The main problem of experimental investigations of local-time effect at the small space distances is resolution enhancement of the macroscopic fluctuations method. As a rule all above-mentioned investigations of local-time effect were carried out by using $\alpha$-decay rate fluctuations of $^{239}$Pu source. But such source of fluctuations becomes uselessness for distances in tens of kilometers or less when histograms duration must be about one second or less. By this reason in the present work we refuse $\alpha$-decay sources of fluctuations. As such a source was chosen noise generated by germanium semiconductor diode. Such source gives noise signal with frequency band up to tens of megahertz and because of this satisfies the requirements of present investigations.

      To check convenience of selected noise source for local-time effect investigations it was tested on distances for which existence of the effect was proved [5]. In cited work was shown appropriateness of semiconductor noise diode for studies of local-time effect.

## 2. Investigation of local-time effect for longitudinal distance between locations of measurements in 15 km.

First series of synchronous measurements were carried out in Pushchino (Lat. 54°50.037′ North, Lon. 37°37.589′ East) and Bolshevik (Lat. 54°54.165′ North, Lon. 37°21.910′ East). Longitudinal difference $\alpha$ between places of measurements is $\alpha = 15.679′$.

This value of α corresponds to difference of local time Δ*t* = 62.7 sec and longitudinal distance Δ*l*, equal Δ*l*=15 km.

Fluctuations from noise generator were digitized with sampling frequency equal to 44100 Hz. In this way in Pushchino and Bolshevik were obtained 10-minute time series. From this initial time series with three different steps (pointed in second line of Table 1) were extracted single measurements and obtained three time series with frequency pointed in third line of Table 1. On the base of this time series in a standard way [1-3] were constructed three sets of histograms. All histograms were constructed using sample length pointed in fourth line of Table 1. Number of histograms per second and durations of histogram for every set are given in the fifth and sixth line of Table 1 correspondingly.

Table 1. Parameters used for calculating sets of 1-, 0.2-, and 0.02-sec histograms.

| 1 | Sampling frequency, Hz | 44100 | 44100 | 44100 |
|---|---|---|---|---|
| 2 | Step, points | 735 | 147 | 14 |
| 3 | Frequency of histogram time series, Hz | 60 | 300 | 3150 |
| 4 | Histogram sample length, points | 60 | 60 | 63 |
| 5 | Histograms per 1 sec | 1 | 5 | 50 |
| 6 | Duration of histogram, sec | 1 | 0.2 | 0.02 |

Fig. 1 presents intervals distribution obtained after expert comparisons of 1-sec histogram sets. The distribution has a peak, which corresponds to time interval equal to 63±1 sec. Taking into account accuracy of synchronization of measurements beginning (0.1 - 0.2 sec) and duration of histograms one can consider this peak to be corresponding with good accuracy to local time difference Δ*t* = 62.7 sec between places of measurements.

Local time peak ordinary obtained on the interval distributions is very sharp and consists of 1-2 histograms [1-3] i.e. is practically structureless. Peak on the Fig. 1 *a*) also can be considered as structureless. This fact leads us to the problem of further investigating of structure of local time peak.

The fact that all sets of histograms were obtained on the base of the same initial time series enables enhancement of time resolution of the method of investigation. Using of 0.2-sec histograms set (forth column of Table 1) increase resolution in five times and allows more detailed investigations of local-time peak structure. Since the positions of the peak on the intervals distribution (Fig. 1) are known it is possible to select their neighborhood by means of 60 sec relative shift of initial time series and prepare after this 0.2-sec histograms set for further expert comparison.

Intervals distribution obtained in result of expert comparisons for 0.2-sec histograms set is presented on Fig. 1*b*). One can see that maximum similarity of histograms shape is observed for pairs of histograms separated by interval in 63±0.2 sec. This value is the same as for 1-sec histograms intervals distribution, but in latter case it is defined with accuracy in 0.2 sec.

It's easy to see from intervals distribution, Fig. 1*b*), that after 5-times enhancement of resolution the distribution has single sharp peak again. So, change of time scale in this case doesn't lead to change of intervals distribution. This means that we must enhance time resolution again to study the local time peak. We can do this by using of 0.02-sec histograms (third line of Table 1).

Intervals distribution for case of 0.002-sec histograms is presented on the Fig. 1*c*). Unlike to intervals distributions on the Fig. 1*a*) and Fig. 1*b*) distribution on the Fig. 1*c*) consists of two distinct peaks. The first peak corresponds to local time difference equal 62.98±0.002 sec, the second one to 63.16±0.002 sec. The difference between the peaks is Δ*t'* = 0.18±0.002 sec.

Splitting of local-time peak on the Fig 1 *c*) is similar to splitting of daily period on two peaks with periods, which equal to solar and sidereal days [6-8]. This fact will be discussed below.

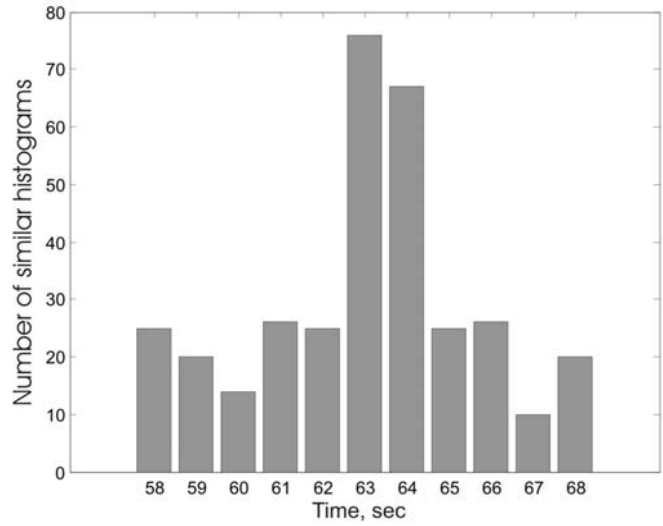

*a*)

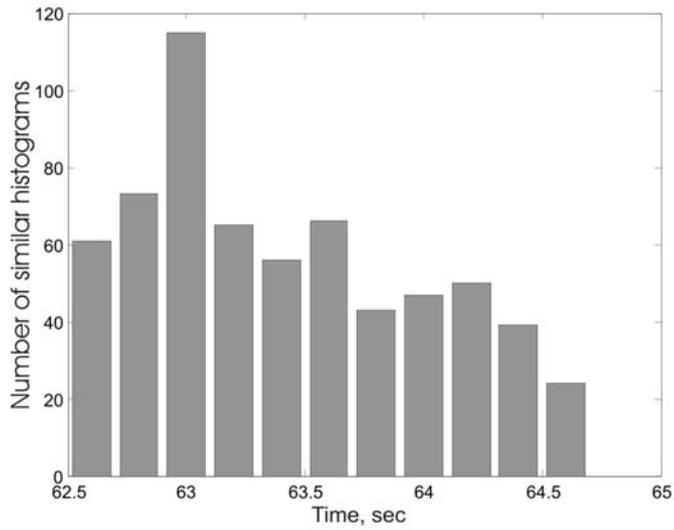

*b*)

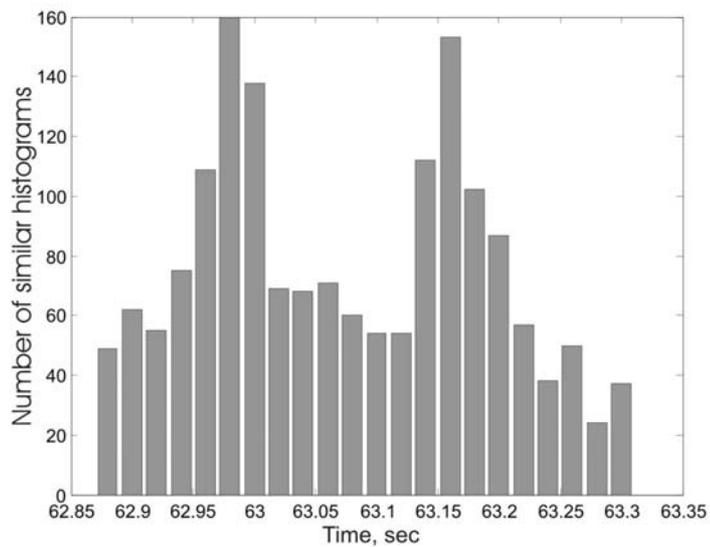

*c*)

Fig. 1. Intervals distributions obtained after expert comparisons of 1-sec (*a*), 0.2-sec (*b*), and 0.002-sec (*c*) histogram sets. Y-axis presents number of histograms, which were found similar; X-axis – time interval between pairs of histograms, sec.

## 3. Investigation of local-time effect for longitudinal distance between locations of measurements from 6 km to 0.5 km. Mobile experiment.

Above presented experiment demonstrates the existence of local-time effect for longitudinal distance between locations of measurements in 15 km and splitting of local-time peak corresponding to the distance. It is natural to investigate the question: which is the minimal distance of local time effect existence? Next step in this direction is the experiment presented below.

Table 2. Locations of mobile measurement system and corresponding parameters.

| № | Locations of mobile measurement system | α | Δ*t*, sec | Δ*l*, km | Δ*t'*, sec |
|---|---|---|---|---|---|
| 1 | Lat. 54°48.16′ N, Lon. 37°43.54′ E | 5.95′ | 23.8 | 6 | 0.066 |
| 2 | Lat. 54°49.28′ N, Lon. 37°41.44′ E | 3.85′ | 15.4 | 3.9 | 0.043 |
| 3 | Lat. 54°50.13′ N, Lon. 37°39.21′ E | 1.618′ | 6.47 | 1.6 | 0.018 |
| 4 | Lat. 54°49.99′ N, Lon. 37°38.13′ E | 0.538′ | 2.152 | 0.5 | 0.006 |

In the experiment two measurement systems were used: stationary with location in Pushchino (Lat. 54°50.037′ N, Lon. 37°37.589′ E) and mobile one. Four series of measurements were carried out. Locations of mobile measurement system for every series of measurements are given in second column of Table 2. Angular longitudinal difference of locations of measurements, α, is presented in third column of the table. Local time difference Δ*t* and longitudinal difference of locations of measurements Δ*l*, are given in fourth and fifth columns of Table 2 correspondingly. Last column gives splitting value of local-time peak, Δ*t'*.

Method of experimental data processing was the same as for experiment presented in second section of the paper. Was found that within accuracy of experiment the local time value Δ*t* and the local-time peak splitting value Δ*t'* can be observed.

## 4. Discussions.

Local-time effect as pointed in [1], is linked to rotatory movement of Earth. The simplest explanation of the fact can be following. Due to the rotatory movement of the Earth after time Δ*t* measurement system No. 2 appears in the same places where was system No. 1. The same places cause the same shape of fine structure of histograms. Actually such explanation is incorrect because of orbital motion of Earth, which noticeably exceeds rotatory movements. Therefore measurement system No. 2 cannot appear in the same places where was system No. 1. But if we consider two directions defined by center of Earth and two points of measurement then after time Δ*t* measurement system No. 2 take the same directions in the space as system No. 1 before. From this it follows that similarity of histograms shapes in some way is connected with the same space directions. This supposition also agrees with experimental results presented in [9-10].

Four-minute splitting of daily period of repetition of histograms shape on solar and stellar sub-periods [3] is the evidence of existence of two preferential directions: to the Sun and to the coelosphere. Really after time interval equal 1436 min the Earth makes one complete revolution and measurement system plane has the same direction in the space as one stellar day before. After four minutes from this moment measurement system plane will be directed to the Sun. This is the cause of solar-day period – 1440 min.

Let us suppose that splitting described in the present paper has the same nature as splitting of daily period. Then from daily period splitting $\Delta T$, which equal $\Delta T = 4\,min$ its possible to obtain proportionality coefficient $k$:

(1) $$k = \frac{240\,\text{sec}}{86400\,\text{sec}} \approx 2.78 \cdot 10^{-3}.$$

Longitudinal difference between places of measurements presented in second section is $\Delta t = 62.7$ sec and we can calculate splitting of local-time peak for this value of $\Delta t$:

(2) $$\Delta t' = k\Delta t = 62.7 \times 2.78 \cdot 10^{-3} \approx 0.17\,\text{sec}.$$

As it is easy to see from Fig. 1c) splitting of local-time peak is equal to 0.18±0.02 sec. This value agrees with estimation (2). Values of splitting of the local-time peak, which are presented in last column of Table 2 also were calculated by help of formula (2). Experiment described in third section shows good agreement of the values with experimentally obtained.

This result allows us to consider sub-peaks of local-time peak as stellar and solar and suppose that in this case the cause of splitting is the same as for daily-period splitting.

Speaking about preferential directions we implicitly supposed that measurement system is directional and because of this can resolve these directions. Such supposition is quit reasonable for the case of daily period splitting but for splitting of local-time peak observed on the small distances becomes problematic since an angle, which must be resolved by the measurement system is neglible. Most likely that in this case we deal with space-time fluctuations, which in some way are connected with preferential directions to the Sun and coelosphire. In other words we can speak about sharp anisotropy of near-earth space-time.

Results obtained in the present work prove possibility of local-time effect investigation on small space scale up to 0.5 km. Farther decreasing of this scale is our immediate task. In the same time sub-peaks obtained as result of splitting of local-time peak also consist of one-two histograms, so are structureless. This fact poses a problem of more detailed investigations of local-time peak structure.

Authors grateful to Dr. Hartmut Muller, V.P. Tikhonov and M.N. Kondrashova for valuable discussions and financial support.